Concept Paper

# Augmented Reality and Gamification: A Framework for Developing Supplementary Learning Tool

Carlo H. Godoy Jr.
Technological University of the Philippines, Philippines
carlo.godoyjr@tup.edu.ph

**Abstract**

*Purpose* – The study's main purpose is to develop a supplementary learning tool framework by the use of a dynamic mobile application using Unity AR and Vuforia for Senior High School (SHS) students and teachers to help the learning process in SHS Earth Science.

*Method* – The researchers will be using the Software Development Life Cycle (SDLC) Model of methodology to ensure the quality of the software as well as the correctness of the development process.

*Results* – The expected result of the study is that Augmented Reality and Gamification will now be used as a supplementary learning tool in SHS Earth Science.

*Conclusion* – Augmented Reality and Gamification can now be used as a supplementary learning tool in SHS Earth Science using the designed framework.

*Recommendations* – Future studies will focus on the development of the framework and the mobile application.

*Research Implications* – Since the system has a lot of potential in the education sector and due to the effects of COVID-19, the software will serve as a pioneer to show that a supplementary tool will help students learn logically and entertainingly especially since schools nowadays are transitioning with either distance learning or blended learning.

## INTRODUCTION

An Augmented Reality or simply called AR is a system that allows an electronically-presented material to be introduced within the interaction of the user and the real world (Carbonell Carrera & Bermejo Asensio, 2017). During the start of the new millennium, people are dreaming of this technology to be available since it was overwhelming how this technology works during that era and no existing technology is comparable to this aside from fiction movies. As that movie introduces the nation to the technology, it became evident that people want this to be a reality and it happened (Peddie, 2017). From that time, the desire of humanity for that technology gave birth to a new phenomenon that nobody knows will be popular for most of the people in the world. Nowadays, the technology emerged on different platforms and now it is being used as a supplement in different industries like healthcare, government, entertainment, and education (Alkhattabi, 2017; Mahmood et al., 2017; Raiff, Fortugno, Scherlis, & Rapoza, 2018; Seevinck, 2017).

The most popular platform of Augmented Reality is a mobile game called Pokémon Go which became a major form of entertainment (Yang & Liu, 2017). Several mobile games nowadays are being used by the education industry as a supplement for learning in different areas of education like English, Language, History, Science, and Mathematics (Atwood-Blaine & Huffman, 2017; Kukulska-Hulme, 2009; Rivadulla, Jesus, & Balahadia, 2017; Tan et al., 2017; Tsai, Cheng, Yeh, & Lin, 2017; ). One of the hardest areas of Mathematics, considered by most students is Precalculus and Calculus since there is a need to apply a lot of critical thinking skills as well as logical reasoning to prove the answer. Some students majoring in Engineering and started to take these subjects tend to either change course or leave the University (Van Dyken & Benson, 2019). In line with this, the student's interest and motivation are being affected by studying, not only that specific subject but also the whole course itself (Van Dyken & Benson, 2019). If the student loses interest and motivation in learning, it will affect not only the teacher but also the rate of education a certain country has. Also, students sometimes are not willing to study because of the enjoyment of playing electronic games (McBride & Derevensky, 2016). These problems may result in poor academic performance, change of mood, and as well as social conflicts.

According to BusinessWorld (2019), it is very evident that there is a decline in the academic performance of students in the Philippines. Mobile game addiction of most Filipinos also poses a great threat in the future of society just like the declining rate of academic performance of the students (Fabito & Yabut, 2018). In a study conducted by the National University of the Philippines, it is identified that mobile game addiction is a concern in the Philippines that grows rapidly which normally increases the risk in the physical as well as the psychological health of a certain individual (Fabito et al., 2018). Based on a report from UNESCO last 2008, the Philippines' Basic Education has low participation and achievement rates because it has fallen dramatically (Wilson Macha, Christopher Mackie, 2018). Mobile game addiction and poor academic performance are among two of the most common problems a student in the Philippines is facing since mobile phones are known to be very affordable nowadays that leads the students to spend more time playing games or using social media than studying, especially if the subject is Science as some of the students think of this as a boring subject. As AR is defined, it is a good avenue to use the mixed reality capability of AR as there are different topics in Earth Science that can't be seen without a 3d representation. This paper aims to turn around the problem of Mobile Game Addiction into a positive one at the same time enhancing the result of their academic performance through an academic game.

*Objectives of the Study*

The general objective of the study is to develop a supplementary learning tool framework by the use of a dynamic mobile application using Unity AR and Vuforia for Senior High School students and teachers to help the learning process in SHS Earth Science.

## REVIEW OF RELATED LITERATURE AND STUDIES

An integrated STEM (Science, Technology, Engineering, and Mathematics) lesson requires participation and nurture students' interest in real-world circumstances. While real-world STEM situations are naturally incorporated, the embedded STEM contents are rarely taught by school educators (Hsu et al., 2017). One of the hardest subjects of that track is Mathematics. One example of a Mathematics subject is Solid Geometry. To give a better experience in learning solid geometry, a study has been conducted to combine Augmented Reality (AR) technology into teaching operations designing a learning scheme that helps junior high school learners learn sound geometry (Liu et al., 2019; TeKolste & Liu, 2018). Based on the result of the study, AR gives a big leap in learning solid geometry. Another study deals with the use of AR in teaching and learning math that uses this technology to its complete benefit in providing concrete experience in interacting with revolutionary solids. At the end of the study, it was found out that Augmented Reality is beneficial in the understanding of computing solids of revolution volumes (Salinas & González-Mendívil, 2017). In this study, Augmented Reality and Gamification are the two solutions that can be replicated by other studies and be transformed into an educational framework.

According to Rouse (as cited in Jung & tom Dieck, 2018) Augmented Reality or simply AR is the integration of information in digital format which includes a live video on the real-time environment of a certain user. In augmentation of live videos, integrating a video picture to digital environment involves identification of an object replicated from the physical world features and will be captured as any format that will be considered as a video picture which will mean that increasing the responsiveness of the generated video picture to the state needed to control the object from the physical world itself (Kochi, Harding, Campbell, Ranyard &Hocking, 2017). In an augmented reality system, the integrated digital information can only be seen using a medium like phone cameras but it will not be seen in the real world. This digital information can be represented in different forms like a stack of virtual cubes or manipulating a non-real object in many ways possible (Hilliges, Kim, Izadi, & Weiss, 2018).

Gamified study or sometimes being called by a lot of people in the academy, Gamification stands for a term used to define how game mechanics are integrated into teaching the process to make instructions more enjoyable and engaging (Fields et al., 2017). It can help learners feel involved in learning through development and progress, recognition and benefits, a greater objective to be pursued, and a sense of teamwork. According to Kapp (as cited by Fields, Lui, & Kafai, 2017), scientists suggest that gamification or gamified learning can be used as an educational instrument to stimulate interest in learning for learners. A lot of games which pretend to have been educational, throughout the manner that as playing, individuals who play them should learn something.

Many electronic games were often intended for instructional reasons. Also, learners using a gamified e-learning platform will generate greater practical test results relative to those using the non-gamified version. Knowledge acquisition was the most frequently reported result for games for learning, while entertainment games addressed a wider variety of affective, behavioral, perceptual, cognitive, and physiological results (Boyle et al., 2016). Learning through games was discovered in a variety of topics with the most common STEM topics and health. A systematic program of experimental job would benefit future studies on digital games, examining in detail which game characteristics are most efficient in fostering participation and learning support (Boyle et al., 2016).

Augmented Reality (AR) apps have received increasing attention over the previous two decades. In the 1990s, AR was first used for apps linked to pilot education as well as for the training of Air Force (Akçayır & Akçayır, 2016). AR generates fresh world experiences with its data layering over 3D space, suggesting that AR should be embraced over the next 2–3 years to give fresh possibilities for teaching, learning, study, or creative investigation according to the 2011 Horizon Report (Chen et al., 2017). AR uses virtual objects or data that overlap physical objects or environments to create a mixed reality in which virtual objects and actual environments coexist in a meaningful manner to increase learning experiences. According to Chiang, Yang & Hwang (as cited by Akçayır & Akçayır, 2016) Augmented Reality is widely used now in the K-12 level in the education industry. Ferrer-Torregrosa et al., (2015) stated that Augmented Reality is also being used now by different universities.

As highlighted in the Horizon report, Augmented Reality (AR) is acknowledged as one of the most significant innovations in greater and K-12 education technology (Johnson et al., 2015). Augmented reality is gradually becoming integrated as an emerging technology in the region of inclusive education that adapts learning on equal footing through exploration and experience by all (Marín-Díaz, 2017). Johnson (as cited by Saltan, 2017) stated that AR is anticipated to be widely adopted in higher education for two to three years and in K-12 for four to five years. It is essential to explore how teachers and scientists incorporate AR into teaching-learning procedures if this is the present state of the art for the use of AR in education. MUVEs and AR became visible in the early 2000s and their effectiveness for learning was soon established by educational research (Dede et al., 2017). Research has shown that AR can be more efficient in supporting teaching than other improved settings in technology. If the content is represented as 3D learners, objects can be manipulated and information handled interactively (El Sayed, Zayed, & Sharawy as cited by Buchner & Zumbach, 2018). Rapid technological evolution has altered the face of education, particularly when technology has been coupled with appropriate pedagogical foundations. This combination has developed fresh possibilities to enhance teaching and learning experience quality (Nincarean et al., 2013).

As a result of discussing the different subjects that Augmented Reality has been used already, it is a great help for students to have an application that has never been created to aid them to have a higher grade. ScavengEarSci will then prove that the difficulty in studying can be easily turned around and the students will start enjoying Earth Science.

## PROPOSED METHODOLOGY

This research uses the Life Cycle of Software Development (SDLC). In each defined stage of the SDLC which will define the software development process, the framework will be described in every part of the process. It comprises a comprehensive plan describing the development of the scheme.

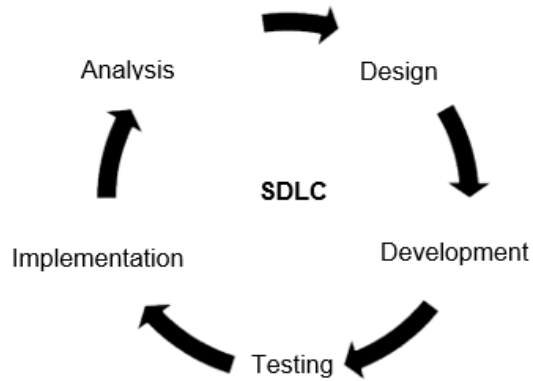

*Figure 2.*Software Development Life Cycle (SDLC) Model

The SDLC comprises several clearly defined stages, namely, Analysis, Design, Development, Test, and Implementation.

A. Analysis
   1. Preparation of questions to be asked to the end-users and technical experts.
   2. Discussion with end-users like students, teachers, and administrators.
   3. Discussion with technical professionals, comprising of developers of Android and IOS apps, to learn some best practices.
   4. Reviewing an existing application that has the same goal, issues, methodology and identifies its weak points
   5. Definition of the process flow of the system using a diagram.
   6. Identification of the needed tools and software that will be needed for the development of the application

B. Design
   1. List the system requirements and characteristics
   2. Build the following: Context Diagram, Data Flow Diagram (DFD), Students Low-Level Flowchart, Teachers Low-Level Flowchart, Use Case Diagram, and Interface Design.
   3. Produce interface design layout and mock-up apps to list possible changes

C. Development
   1. Use Vuforia and Unity to convert the final design layout to a working interface.
   2. Use MySQL to create the database and store processes.
   3. Use C # to create the API (Application Programming Interface) for the key system functionality.
   4. Use Vuforia and Unity to integrate the interface and API to create one working application.

D. Testing
   1. Identification of tiny components of apps that are testable called units.
   2. Test for the correct operation of each unit.
   3. Test the system's general integrity.
   4. Test the system based on functionality, reliability, and compatibility using test instances and live testing with actual end-users.

E. Implementation
   Request the application to be in Playstore and Apple Appstore.

## *Project Design*

ScavengEarSci will be designed similar to Pokemon-Go but with a twist of the traditional scavenger hunt game. This mobile game will be investigatory wherein the learners will learn the module from beginning to end while solving the mystery of the game. If the mobile game will be designed in a somehow almost perfect manner, a game-based learning approach for teaching Earth Science will control and make use of the student's intrinsic motivation together with their interest in playing and lead them in a very interactive way of solving exercises and seat works for the Earth Science subject.

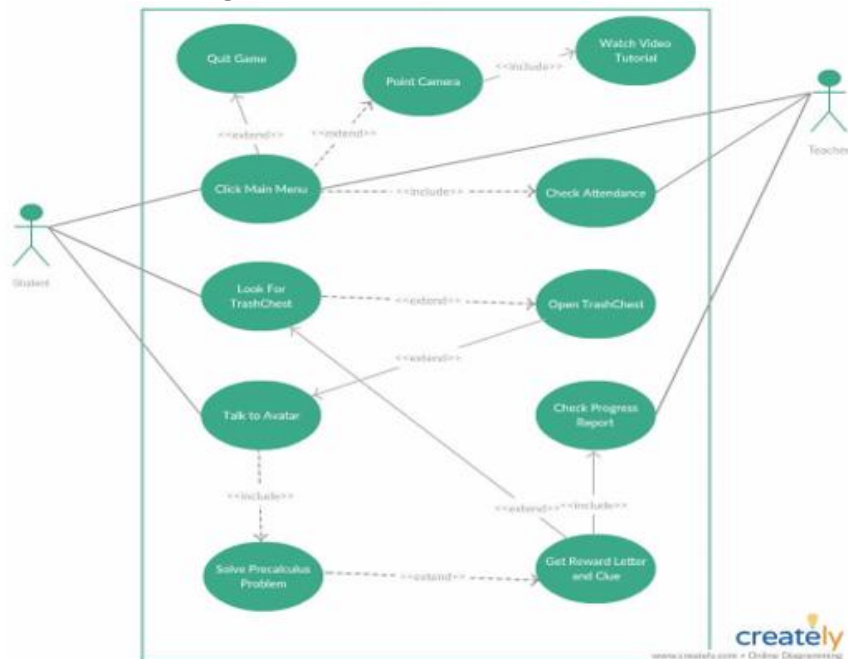

*Figure 3.* Use Case Diagram of ScavengEarSci

Figure 3 shows the Use Case Diagram of the application which is named ScavengEarSci. This also illustrates and describes the interaction of the users with the system. The system has two types of users namely Student and Teacher. Both the teacher and the student will have the ability to click on the main menu. Once on the main menu, the teacher can go ahead and check the attendance and the progress report of the students. On the other hand, the student will have two options after clicking on the main menu.

The first option of the user is the tutorial mode. In the tutorial mode, the students can point out the camera in the physical environment. Once the camera has detected any symbol or object related to any topic in Earth Science, it will go ahead and play the tutorial video for that certain topic. The second option for the student is to play the ScavengEarSci Mystery Game. The student will walk around the area and look for a TrashChest. Once the student found a TrashChest the option will be to open it or not. If the student chooses not to open it, the student will just look for another TrashChest. If the student will choose to open it, an Earth Science problem and Tutorial will appear and the student will need to solve the problem after the tutorial. If the student got the answer wrong, the student will need to walk around again looking for another TrashChest. If the student got the answer right, the student will get another Reward letter and will have the chance to go on a quest to complete the Mystery word following the clue given by the Avatar. All of the points garnered will be saved to the progress report afterward.

Figure 4 shows the Flowchart of the ScavengEarSci application for both the teachers and the students. It illustrates an overview of the overall process flow of the system. The system's process starts after verifying if the user is a student or a teacher. Once verified, the user when then starts clicking the main menu. If the user is a teacher, the option will prompt the progress report as well as the attendance report of the students profiled under the teacher. Once done it will give the teacher the option to either quit or go back to the main menu. On the other hand, if the user is a student and the main menu has been clicked, three options will appear. The three options are Tutorial Mode, Scavenger Hunt, and Reports. The student will have the option to play the mystery game or just trigger a tutorial by augmenting an image else the use chose the progress report. If the tutorial mode has been chosen, the system will try to recognize a shape in the physical world using a shape detection algorithm, once a shape has been detected it will then compare if the shape is in the multi-media database. If there is a match, the video tutorial will play else it will go back to the shape detection algorithm to find another shape.

If the user selects the Scavenger Hunt Mode, the system will find the location of a trash chest, once the location match, the user will be given an option to either open it or not. If the user decided not to open it, the system will let the user walk and look for another trash chest. If the user decided to open it an avatar will appear to ask the user to solve a problem. If the user decided to answer it, a reward letter and clue will be given if incorrect it will ask the user to answer the problem again. The system will loop until all the topics have been completed. Once the topic is completed, it will give the user an option to either quit the application or go back to the main menu. The third option for the user is to check his attendance and progress report. Once done it will give the student the option to either quit or go back to the main menu.

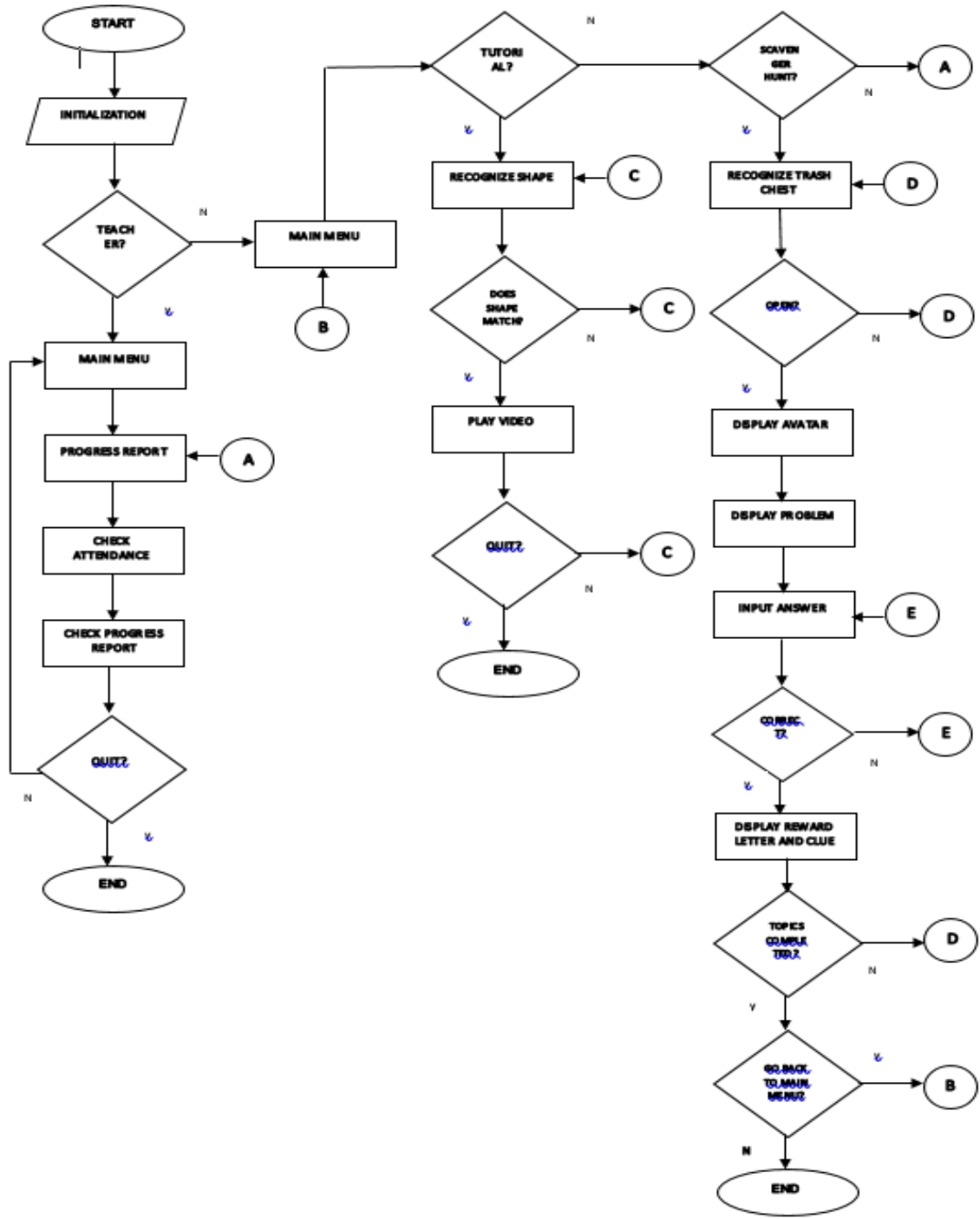

*Figure 4.* System Flowchart

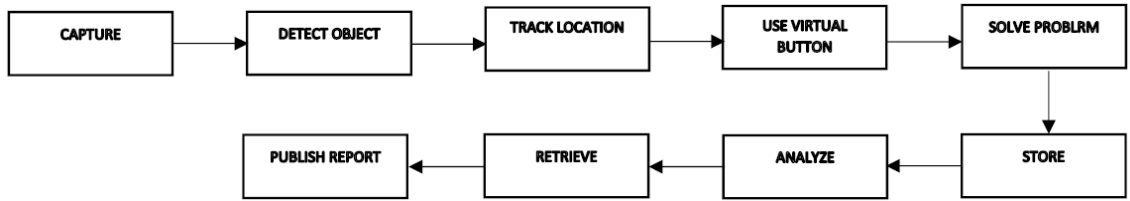

*Figure 5.* Architecture Design of Location-Based Tracking Algorithm

Figure 5 illustrates the Architecture Design of the Location-Based Tracking Algorithm. The process will start when the user pointed out the camera to the specified location while walking. Then, when the camera pointed out the location where the TrashChest is located, it will prompt the user to open it or not. If the user decides to open the TrashChest and solve the problem with a correct answer, the score will then be generated as data to be stored in the Cloud Database. When either the student or the teacher will check the generated result, the data analysis tool will then retrieve the data from the cloud database and generate a report from the retrieved data.

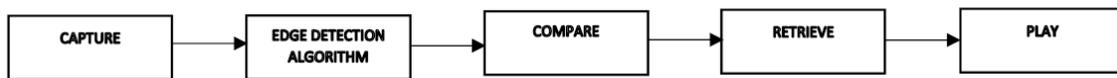

*Figure 6.* Architecture Design of Image Recognition Algorithm

Figure 6 shows the Architecture Design of the Image Recognition Algorithm. The process will start when the user points out the camera on the specified image. The image recognition algorithm will then read the image. The image will be check if it's a circle, parabola, hyperbola, or ellipse. Once recognized, the image will then be evaluated. After evaluation, it will be compared to the videos stored in the multimedia database. Once the video related to the image has been identified, the video will then be retrieved and it will start playing as a tutorial video.

Figure 7 describes the procedure in developing the Augmented Reality Application. To start the development of the application, the Vuforia program will be used to create the database. This database will contain images with different goals that can be included in the Picture Target to be detected and identified by the use of an AR camera. Vuforia will also be utilized to produce the activation key, which will be then subsequently used in Unity. The application key will be unique for each project. The Unity software will then export the objects from the database created in Vuforia and will be included in the prefabs and which can be a picture target where the digital picture elements will then be deployed. For the animation, there are models in Unity that we can select. After selecting the 3d model, specifying where and when to run the animation must be done. It can be done by using a programming language that can be used in Unity. The programming language is C#, which is very oriented to objects. By using this, inserting the animation for the application will be easier. Audios will then be introduced to supplement. For this process, using an algorithm that normally detects an image while playing a sound corresponding to it will be used. After that is the development of the application for IOS and will make sure that it will be compatible with the Android operating system as well. It will be done in Unity by making sure that the application name will be very unique, a characteristic image will need to be created which will represent the application.

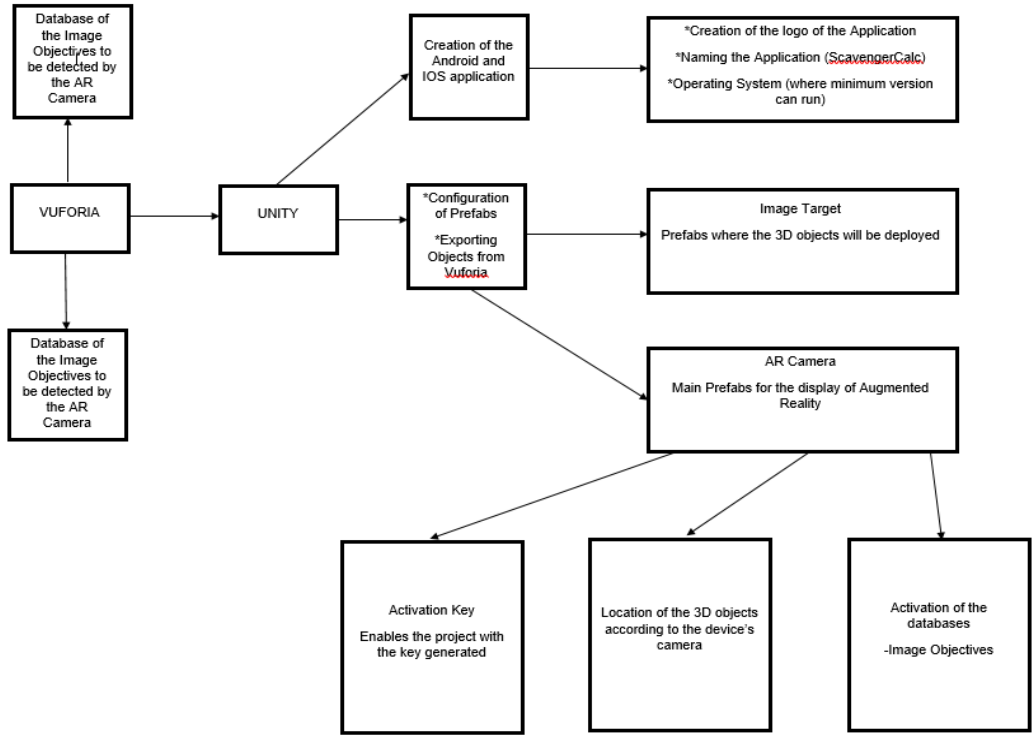

*Figure 7.* Summary of the Development of ScavengEarSci

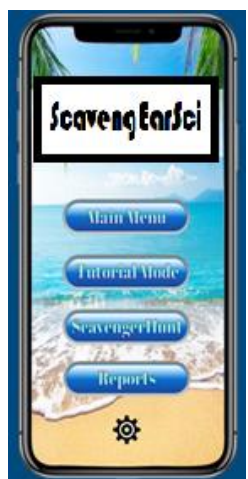

*Figure 8.* Proposed Main Menu

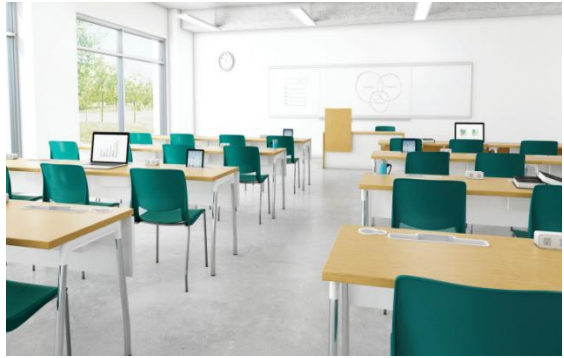
*Figure 9.* Physical Environment

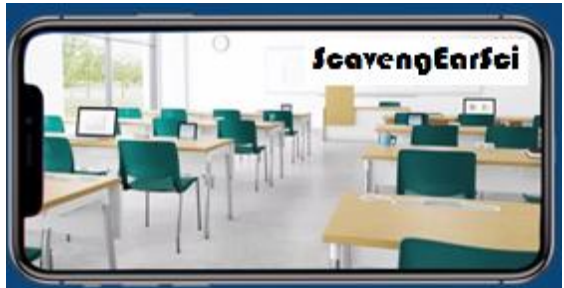
*Figure 10.* Physical Environment when pointed a Camera

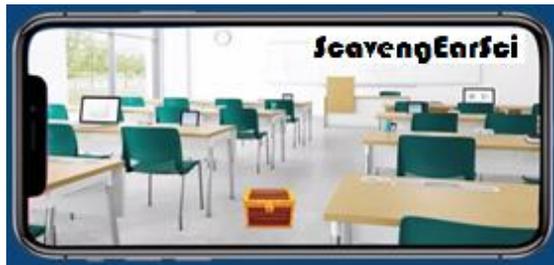
*Figure 11.* Physical Environment when pointed a Camera through Scavengercalc

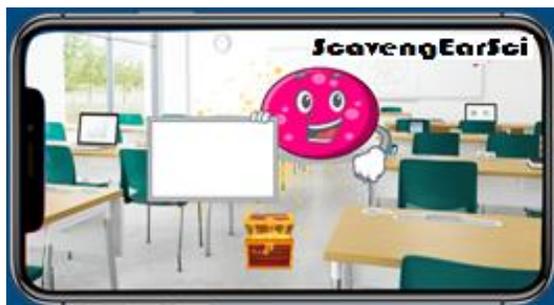
*Figure 12.* Scenario 3 when the player opened the TrashChest

Figure 8 to 12 shows the Graphical UserInterface Design of ScavengerCalc. The application will initially show the main menu, where the user can either choose the Tutorial Mode, Scavenger hunt, or Reports. If the user will click the Tutorial Mode, it will show the camera then it will look for an object like any topic under Earth Science. Once an image has been recognized by the Image Recognition algorithm, it will look for a similar topic in the database then it will play the video tutorial. If the user will click the Scavenger Hunt under the main menu, it will show the AR Camera and the user will need to work around the area looking for a TrashChest.

Once the student found a TrashChest the option would be to open it or not. If the student will choose not to open it, the student will just look for another TrashChest. If the student chooses to open it, an Earth Science problem would appear with a funny avatar and the student would need to solve the problem. If the student got the answer wrong, the student would need to walk around again looking for another TrashChest. If the student got the answer right, the student would get the Reward letter and will go on a quest to complete the Mystery word while following the clue given by the Avatar. All of the points garnered will be saved to the progress report afterward. On the other hand, if the user would click on reports, the attendance and the progress score of the student will be seen.

## PROPOSED OPERATION AND TESTING PROCEDURE

To ensure system quality, a series of tests will be conducted for each module and by installing the application in a live environment, the system will be subjected to real-life testing.

**Functionality Test.** The functionality test will be used to guarantee that the application meets all the criteria and implements all the functionalities indicated in its functional requirements. Following the creation stage, the investigator will take the following measures:

1. Prepared test cases for functionality for each module.
2. The test cases will be executed.
3. Will record the output of the test.
4. The failed test cases will be analyzed and remedied.
5. Failed test instances will have re-executed to check that the test instances have been remedied.

The test cases will include a set of inputs, execution preconditions, and the expected results. The test case form that will be used is illustrated in Table 1.

**Reliability Test.** This test will be performed to confirm that for a particular period the system can perform the designated tasks correctly. Live testing will be performed to check the system's reliability. The system will be implemented and exposed to real-world testing in a live setting. The following measures are to be taken:

1. Live deployment of the scheme will be initiated.
2. Will request possible end-users which will include IT experts, students, teachers, and possible administrators to install, use, and to test the application for a certain period.

3. Will ask end-users to forward back to the investigator the problems experienced.
4. Will collect the outcomes and tabulate them.

The investigator will also perform the following measures concerning a live application for reliability testing:

1. Preparation of test cases for reliability test.
2. Will perform every test case.
3. Records the output of every test case.

Any reliable test case will be documented using a test case form. The found problems will be documented in a test incident log and a summary of the execution of the reliability test case will be collated. The total number of test cases will be executed and will be summarized in a table, to check whether the application is reliable under different instances.

**Portability Test**. Portability test will be used to explore the efficiency and effectiveness of the system with distinct devices that have distinct requirements such as variants of Android OS and IOS, hardware settings, screen sizes, and kinds or speeds of the network. The experiments will be provided in Table 8 to install and use the application in multiple kinds of settings. If all the anticipated behaviors are met, the testing is effective. The compatibility tests are provided in the Appendix for each module.

## PROPOSED EVALUATION PROCEDURE

The following actions will be carried out to evaluate the application performance and the study results:
1. Invite panels and evaluators, which will include 10 IT experts and 5 Senior High school students, and 5 teachers.
2. The application will be downloaded and installed.
3. The application will be used after downloading and installation.
4. Will Request the panels and evaluators to rate the application using the ISO 25010 Evaluation Criteria for Software Materials and the rating scale would range from 1-5 where 5 will serve as the highest and 1 will serve as the lowest.
5. Collect and tabulate the information to obtain the general mean and mean for each criterion.

Use the interpretation to interpret the numerical value to obtain the corresponding descriptive score.

Table 1. Portability Testing Summary

| Environment | Expected Behavior |
|---|---|
| **Android OS Versions** | |
| 5.0 | The user has no problem accessing all the modules |
| 6.0 | The user has no problem accessing all the modules |
| 7.0 | The user has no problem accessing all the modules |
| 8.0 | The user has no problem accessing all the modules |
| **IOS Versions** | |
| 8 | The user has no problem accessing all the modules |
| 9 | The user has no problem accessing all the modules |
| 10 | The user has no problem accessing all the modules |
| 11 | The user has no problem accessing all the modules |
| **Network** | |
| 2G | The user has no problem accessing all the modules |
| 3G | The user has no problem accessing all the modules |
| 4G (LTE) | The user has no problem accessing all the modules |
| **Screen Sizes** | |
| 426dp x 320dp | The user has no problem accessing all the modules |
| 470dp x 320dp | The user has no problem accessing all the modules |
| 640dp x 480dp | The user has no problem accessing all the modules |
| 960dp x 720dp | The user has no problem accessing all the modules |
| **Hardware Configurations** | |
| *RAM* | |
| 1GB | The user has no problem accessing all the modules |
| 2GB | The user has no problem accessing all the modules |
| *CPU* | |
| Dual-core | The user has no problem accessing all the modules |
| Quad-core | The user has no problem accessing all the modules |
| Octa-core | The user has no problem accessing all the modules |

## CONCLUSION AND FUTURE WORKS

Game-based learning as proven by different researches improves the learning experiences of the student inside the classroom. This kind of approach provides social, emotional, and cognitive development for the students. The use of Augmented Reality based games on the other hand has been proven to provide a partial solution to the declining rate of student's motivation and engagement in different learning areas in the Education Sector. The integration of new methods in teaching like gamification and game based-learning approaches can help the students to have a higher grade in different subjects.

For future studies, an Augmented Reality Based mobile application named ScavengEarSci will be developed by the researcher. The proposed conceptual framework and methodology discussed in this study will be used in the development to ensure the quality and correctness of the process. Once the mobile application has been developed, it must be evaluated by IT experts using ISO 25010. Based on their future assessments the application will be improved if there is a need to. With this, it can be assessed that the application can be a good supplementary tool in improving the learning experiences of Senior High school students that have Earth Science.

## ACKNOWLEDGEMENT

The author would like to thank the continued support of the management of the Technological University of the Philippines – Manila especially from the College of Science and the College of Industrial Technology, particularly in their continued passion to produce competent researchers in this field. Likewise, gratitude is given to the following people who provided financial support for the publication of this paper to support research about new normal education – namely: Danica Rose Tiozen, Lester Santos, Emilyn Saavedra, Samuel John Cruz, Jam Quismundo, Bellie Jay Lasac, Jerome Napalang, Judith Salas Pascua, Maria Victoria Jazareno, Reese Umali, Reynaldo Juico, and Carlo Ryoichi Sakuma.

**Author's Biography**

Mr. Carlo H. Godoy Jr is certified Fortinet's Associate Network Security Engineer (NSE 1 & 2). He is also a Support Analyst for SQL at Human Edge Software Philippines. He is a former Escalations Manager at Novartis Pharmaceutical. A Research Scholar and a graduating masters in information technology student at the Technological University of the Philippines specializing in studies about Emerging Technologies. He has few research projects about Augmented Reality and Microprocessor-Based Systems.

ORCID ID:  HYPERLINK "https://orcid.org/0000-0002-7701-8036" https://orcid.org/0000-0002-7701-8036

Web of Science ResearcherID:  HYPERLINK "https://publons.com/researcher/AAO-2785-2020/" \o "Copy and share this profile's URL" AAO-2785-2020